\documentclass[12pt]{iopart}

\usepackage{graphicx,epsfig}
\usepackage{color}
\usepackage[latin1]{inputenc}

\newcommand{\wmk}{Wm$^{-1}$K$^{-1}$}

\begin{document}

\title{Thermal conductivity of one-, two- and three-dimensional carbon}

\author{Luiz Felipe C. Pereira$^1$, Ivana Savi\'c$^2$ and Davide Donadio$^1$}
\address{$^1$Max Planck Institute for Polymer Research, Ackermannweg 10, D-55128 Mainz, Germany.}
\address{$^2$Tyndall National Institute, Lee Maltings, Dyke Parade, Cork, Ireland.}
\ead{pereira@mpip-mainz.mpg.de, donadio@mpip-mainz.mpg.de}

\date{\today}

\begin{abstract}

Carbon atoms can form structures in one, two, and three dimensions due to its unique chemical versatility.
In terms of thermal conductivity, carbon polymorphs cover a wide range from very low values with amorphous carbon to very high values with diamond, carbon nanotubes and graphene.
Schwarzites are a class of three-dimensional fully covalent sp$^2$-bonded carbon polymorphs, with the same local chemical environment as graphene and carbon nanotubes, but negative Gaussian curvature. 
We calculate the thermal conductivity of a (10,0) carbon nanotube, graphene, and two schwarzites with different curvature, by molecular dynamics simulations based on the Tersoff empirical potential.
We find that schwarzites present a thermal conductivity two orders of magnitude smaller than nanotubes and graphene.
The reason for such large difference is explained by anharmonic lattice dynamics calculations, which show that phonon group velocities and mean free paths are much smaller in schwarzites than in nanotubes and graphene.
Their reduced thermal conductivity, in addition to tunable electronic properties, indicate that schwarzites could pave the way towards all-carbon thermoelectric technology with high conversion efficiency. 

\end{abstract}



\submitto{\NJP}

\maketitle

\section{Introduction}

Carbon is arguably the most versatile element of all. Besides its importance in organic matter and life itself, carbon also assembles in structures covering the whole range of dimensionalities accessible in the real world.
Up to a few decades ago only three carbon allotropes were know: graphite, diamond and amorphous carbon. The discovery of fullerenes expanded the list to include buckyballs~\cite{Kroto1985} and carbon nanotubes~\cite{Iijima1991}. More recently, the isolation of individual graphite sheets provided the family of carbon allotropes with yet another component named graphene~\cite{Novoselov2004}. 
While buckyballs are cage-like carbon molecules, i.e. 0D structures, and carbon nanotubes (CNTs) extend in one dimension (1D), atomically thin graphene takes the form of 2D membranes. 
Both nanotubes and graphene can be grown to reach sizes beyond the micrometer scale.   
Three-dimensional (3D) fully covalently bonded sp$^2$ carbon forms were predicted more than  two decades ago~\cite{Mackay:1991tl,Lenosky:1992ug,Townsend:1992vx,Gaito:1998ur,Terrones:2003uz}, and were eventually grown, although in a non-periodic form, by supersonic cluster beam deposition~\cite{Donadio:1999ur,Barborini:2002dl}. 
The periodic structures proposed in the earlier theoretical works correspond to a tessellation with hexagons and larger polygons, usually heptagons and octagons, of the negatively curved minimal surface proposed by the mathematician K.~H.~A. Schwarz in 1890.

Physical properties of carbon-based materials, in particular nanotubes and graphene, have been under intense investigation for the last two decades. 
Light atomic weight and strong covalent bonding provide carbon materials with special mechanical and thermal properties, which make them appealing for technological applications~\cite{Balandin2011}.
For example, diamond has the highest thermal conductivity ($\kappa$) among bulk insulators. 
Graphite has a high in-plane $\kappa$ and a very small cross-plane one. In the case of graphite this anisotropy is due to the presence of strong covalent bonds between atoms in the same plane, but only a weak Van der Waals interaction between individual planes.
CNTs and graphene have come to rival diamond presenting even higher thermal conductivities. 
These nanostructures display very similar chemical bonding features, where carbon atoms are connected by strong sp$^2$ bonds.
The same type of strong sp$^2$ bonds characterize carbon schwarzites, which in principle could also have high thermal conductivities.
However, schwarzites are also porous materials with very low density, bearing a resemblance with nano-phononic crystals, which then would favour a low $\kappa$.
Indeed, nano-porosity and low density have been shown to lower the heat transport capabilities of materials~\cite{Yu:2010fp,He:2011fh}.
Therefore, the actual $\kappa$ of either crystalline or random schwazites is difficult to predict on the basis of general considerations and remains unknown. 

Because of their high thermal conductivites carbon nanostructures have been regarded as potential building blocks for thermal management and passive cooling devices~\cite{Pop:2010dv}. 
In contrast, routes to reduce $\kappa$ of carbon nanostructures and improve their thermoelectric figure of merit have also been explored, laying the basis for carbon-based thermoelectric technology. 
Theoretical studies suggest that a significant reduction of $\kappa$ with respect to pristine CNTs and graphene can be achieved in anti-dotted graphene lattice~\cite{Gunst2011}, graphene nanoribbons with disordered edges~\cite{Sevincli:2010cc} or a specific edge design~\cite{Sevincli:2013hr}, and chemically functionalized graphene~\cite{Kim:2012gn}. Although promising, these approaches rely on the use of atomically thin nanostructures, which would limit the range of processing and application. It is then worth investigating whether carbon schwarzites may respond to the need for low-$\kappa$ 3D carbon.

\begin{figure}[htbp]
\hfill
\includegraphics[width=0.75\linewidth]{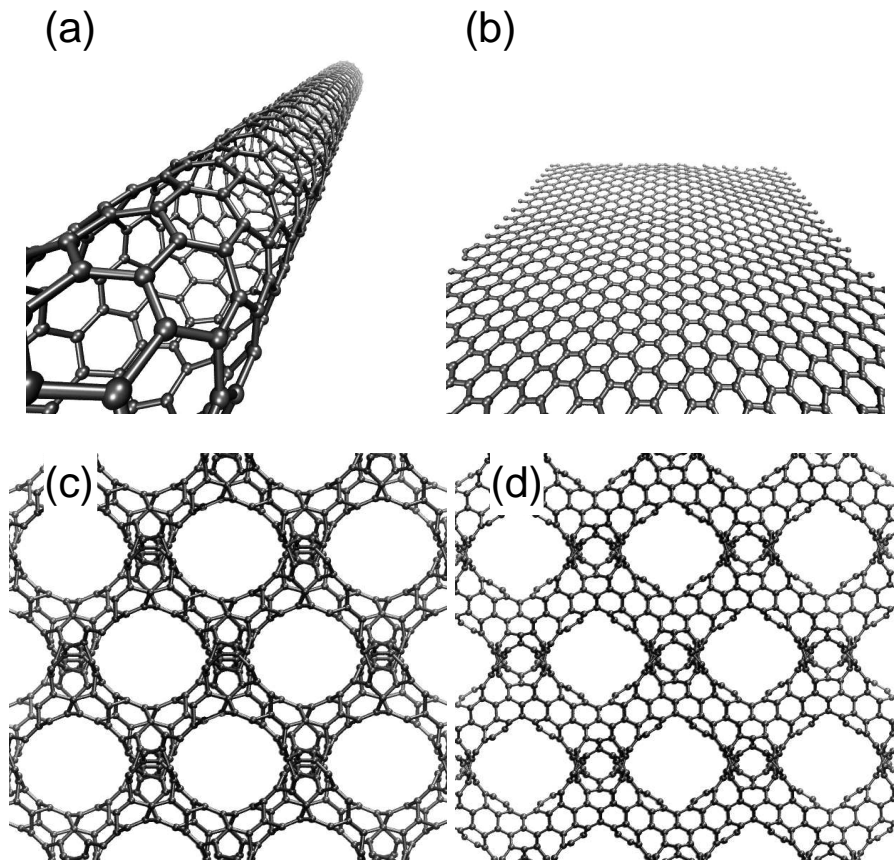}
\caption{Beads and sticks representation of a (10,0) carbon nanotube (1D, panel a), graphene (2D, panel b), and of two $fcc$ schwarzites (3D), namely (C$_{28})_2$ (c) and (C$_{64})_2$ (d), both replicated in space and seen from the (110) direction. The different Gaussian curvature between the two schwarzitic structures appears evident from radii of the connections between the tetrahedral junctions.}  
\label{fig:structures}
\end{figure}

In this work we  investigate thermal transport  in 1D, 2D, and 3D sp$^2$ carbon materials, computing the thermal conductivity of a single wall carbon nanotube (SWCNT), suspended graphene sheets and two crystalline $fcc$ schwarzites (figure \ref{fig:structures}) by extensive molecular dynamics (MD) simulations. 
Whereas the thermal conductivity of graphene and SWCNT is very high, on the order of 10$^3$  \wmk, we find that schwarzites have a thermal conductivity on the order of 10  \wmk, which is surprisingly low for a crystalline carbon polymorph. 
We use anharmonic lattice dynamics (LD) to unravel the reason behind such large differences in thermal conductivity among carbon nanostructures with different dimensionality.

The remaining is organized as follows: in the next section we review the equilibrium MD method to compute thermal conductivity, the LD calculations, and the empirical interatomic potential used to describe the covalent carbon-carbon bonds. 
In section~\ref{sec:results} we report the results of our thermal conductivity calculations by MD, including detailed size convergence studies, which are critical in this type of calculations. 
An interpretation of the MD results is provided in section~\ref{sec:LD}, in which the phonon properties of the systems are comparatively analysed via anharmonic LD. 
The main results are then summarized in section~\ref{sec:conclusions}.

\section{Methods}
\label{sec:methods}

\subsection{Green-Kubo Thermal Conductivity}

Linear transport coefficients can be calculated from equilibrium fluctuations of the respective conjugate flux via Green-Kubo relations~\cite{Green1954, Kubo1957}.
In the case of heat transport, $\kappa$ is calculated from fluctuations of the heat flux via the heat flux autocorrelation function (HFACF).
For a system in equilibrium (in the absence of a temperature gradient) the heat flux averages to zero over time, but the HFACF decays with time and its decay time is proportional to the lifetime of the heat carriers (phonons).
In the present work we calculate phononic thermal conductivity via atomistic equilibrium MD simulations, where the heat flux is calculated from atomic quantities such as velocity, energy and virial atomic stress.

The Green-Kubo expression for each individual component of the thermal conductivity tensor can be written as
\begin{equation}
\kappa_{\alpha \beta} =  \frac{1}{k_BT^2}\lim_{t\to\infty} \lim_{V\to\infty} \frac{1}{V} \int_0^t \langle J_{\alpha}(t') J_{\beta}(0) \rangle dt',
\label{eq:greenkubo}
\end{equation}
where $\mathbf{J}$ is the heat flux, $k_B$ is Boltzmann's constant, $T$ is the system temperature and $V$ its volume.
The HFACF should decay to zero for $t$ longer than the correlation time of the system, and the respective integral saturates at a constant value. In practice, at long times the HFACF becomes noisy and the integral diverges because of this statistical noise. 
Therefore, $\kappa_{\alpha \beta}$ is taken as the stationary value of  (\ref{eq:greenkubo}) before it drifts due to the accumulated noise.
Finite size effects can have a large influence on the fluctuations of physical quantities during MD simulations, therefore it is necessary to ensure  convergence of $\kappa$ with system size as represented by the limit to large volume in (\ref{eq:greenkubo}).

In general, $\kappa$ is a second-order tensor with six independent components. However,  symmetry can be applied to reduce the number of independent components for specific systems.
In the case of 1D materials such as a SWCNT, the system can be oriented such that there is  only one non-vanishing component, along its axis, $\kappa_{zz}$.
In the case of graphene, because of its hexagonal 2D symmetry, there are only two independent but equivalent non-vanishing components $\kappa_{xx}=\kappa_{yy}$. 
For 3D materials with cubic symmetry, such as the fcc schwarzites here considered, there are three equivalent non-zero components $\kappa_{xx}=\kappa_{yy}=\kappa_{zz}$.

Finally, applying (\ref{eq:greenkubo}) to low-dimensional systems such as nanotubes and graphene requires a suitable definition of volume. 
Here we assign a nominal thickness of $h=0.335$ nm to a sheet of graphene, such that its volume is given by the product of this thickness with the area of the sheet, $V_{graphene}=L_x L_y h$. 
Similarly, the volume of a SWCNT is approximated to that of a cylindrical shell with an inner radius $r_- = r-h/2$ and outer radius $r_+ = r+h/2$, where the radius of a $(m,0)$ SWCNT is defined as $r=a_{CC} \sqrt{3}m/ 2 \pi$ and $a_{CC}=0.144$ nm is the carbon-carbon distance. In this way we can write $V_{SWCNT}=\pi (r_+^2 - r_-^2) L_z$.

\subsection{Anharmonic lattice dynamics and the Boltzmann-Peierls equation}

In the harmonic approximation of LD the dynamical matrix $\mathbf{D}$ is defined as:
\begin{equation}
D_{ij}(\mathbf{q})=\frac{1}{\sqrt{m_i m_j}}\frac{\partial^2 V}{\partial x_i \partial x_j}\exp(i\mathbf{r} _{ij}\cdot\mathbf{q})
\label{dynmat}
\end{equation}
where $V$ is the potential energy of the system at equilibrium, $m$ indicates atomic masses, $\mathbf{r}$ the distances between pairs of atoms, and the indexes $i$ and $j$ group the atomic and the Cartesian indexes.
In a system of $N$ particles $\mathbf{D}$ is then a $3N\times 3N$ matrix.
Solving the eigenvalue problem:
\begin{equation}
\mathbf{D}(\mathbf{q})\mathbf{e}_\lambda(\mathbf{q})=\omega^2_\lambda(\mathbf{q})\mathbf{e}_\lambda(\mathbf{q})
\label{eig}
\end{equation}
provides the frequencies $\omega_\lambda(\mathbf{q})$ and the normalized displacement vectors $\mathbf{e}_\lambda(\mathbf{q})$ of the normal modes of the system.  

To compute the phonon lifetimes, one has to consider phonon-phonon scattering processes (normal and Umklapp). Generally the main scattering contribution comes from 3-phonon scattering processes, in which either two phonons ($\omega_\lambda$,$\omega_\mu$) annihilate into a third phonon ($\omega_\nu)$, or one phonon ($\omega_\lambda$) gives rise to two phonons ($\omega_\mu$,$\omega_\nu$). Energy and momentum conservation determine the following selection rules for the two processes:
\begin{eqnarray}
 \omega_\lambda(\bf{q}) \pm \omega_\mu(\bf{q}^\prime) - \omega_\nu(\bf{q}^{\prime \prime})  = 0 \\
 \bf{q} \pm \mathbf{q}^\prime -  \mathbf{q}^{\prime \prime} = \mathbf{Q}
 \label{selection}
\end{eqnarray}
where $\mathbf{Q}$ is a reciprocal lattice vector. 
The lifetime $\tau_\lambda(q)$ of phonon $\lambda$ at q-point $\mathbf{q}$ is given by \cite{Maradudin:1968ve, Turney:2009bb}:
\begin{eqnarray}
 \nonumber 1/\tau_\lambda({\bf q}) =\frac{\hbar\pi}{4N_q\omega_\lambda(\bf{q})} \sum_{\mu q^\prime \nu q^{\prime\prime}}\frac{|V_{\lambda\mu\nu}(\bf{q},\bf{q}^\prime,\bf{q}^{\prime\prime})|^2}{\omega_\mu(\bf{q}^\prime)\omega_\nu({\bf q}^{\prime\prime})} \delta_{{\bf Q,q\pm q^\prime-q^{\prime\prime}}} \cdot \\
 \nonumber  \Big [0.5(1+n_\mu({\bf q}^\prime)+n_\nu({\bf q}^{\prime\prime}))\delta(\omega_\lambda({\bf q})-\omega_\mu({\bf q}^\prime)-\omega_\nu({\bf q}^{\prime\prime})) \cdot \\
   + (n_\mu({\bf q}^\prime) - n_\nu(\bf{q}^{\prime\prime}))\delta(\omega_\lambda(\bf{q})-\omega_\mu(\bf{q}^\prime)+\omega_\nu(\bf{q}^{\prime\prime}))\Big ]  
\label{Gamma}
\end{eqnarray}
where $V_{\lambda\mu\nu}$ are the coupling constants between phonons $\lambda,\mu$ and $\nu$, and $n$ are the equilibrium phonon populations, which can be chosen according to either quantum (Bose-Einstein, BE) or classical statistics. 
Although the correct statistics for phonons is BE,  it is possible to use classical occupancies to estimate the entity of quantum effects and to compare to classical  molecular dynamics simulations, which necessarily obey classical statistics ~\cite{He2012a}.
$V_{\lambda\mu\nu}$ are the third-order anharmonic coefficients of the interatomic potential $\mathcal V$ projected on the normal mode coordinates:
\begin{equation}
V_{\lambda\mu\nu}({\bf q,q^\prime,q^{\prime\prime}}) = \sum_{i,j,k} \frac{\partial^3 {\mathcal V}}{\partial x_i\partial x_j\partial x_k}\frac{e^\lambda_i({\bf q})}{\sqrt m_i}\frac{e^\mu_j({\bf q^\prime})}{\sqrt m_j}\frac{e^\nu_k({\bf q^{\prime\prime}})}{\sqrt m_k}\exp(i\mathbf{r} _{i}\cdot\mathbf{q})\exp(i\mathbf{r} _{j}\cdot\mathbf{q^\prime})\exp(i\mathbf{r} _{k}\cdot\mathbf{q^{\prime\prime}})
\end{equation}
where  $i,j,k$ express Cartesian indexes and $r_{i,j,k}$ are atomic coordinates as in (\ref{dynmat}).

Phonon propagation is described by the Boltzmann transport equation (BTE):
\begin{equation}
 \frac{\partial n_\lambda}{\partial t} + \mathbf{v}_\lambda\cdot \mathbf{\nabla} n_\lambda = \left( \frac{dn_\lambda}{dt}\right)_{scat}.
\label{boltz}
\end{equation}
The righthand side of (\ref{boltz}) accounts for the scattering processes that create or destroy phonons. These include the already mentioned third order anharmonic scattering processes, but also higher order terms, as well as impurity and boundary scattering. 
The non-equilibrium occupation function can be written as $n_\lambda = n^0_\lambda+\delta n$, and,
assuming small temperature gradients ($\nabla T$), one can linearize  (\ref{boltz}) by treating $\delta n$ perturbatively. In stationary conditions the first term of (\ref{boltz}) vanishes, and the linearized BTE is written in the form
\begin{equation}
  \mathbf{v}_\lambda\cdot \nabla T  \frac{\partial n^0_\lambda}{\partial T} = \left( \frac{dn_\lambda}{dt}\right)_{scat}.
\label{lBTE}
\end{equation}
This linearized BTE can be solved at different levels of accuracy and complexity.
The simplest approach is the relaxation time approximation (RTA). By this approach one calculates the relaxation time of each phonon, assuming that all the other modes have the equilibrium occupation ($n^0_\lambda$). 

The resulting expression for $\kappa$ is the sum over the contribution of all independent phonon modes
\begin{equation}
  \kappa = \sum_\lambda \kappa_\lambda = \sum_\lambda C_\lambda v^2_\lambda\tau_\lambda 
  \label{bte-smrt}
\end{equation}
where $C_\lambda$ is the specific heat per unit volume of each vibrational state.
By making use of anharmonic LD as outlined above, one can determine the anharmonic phonon lifetimes, usually limited to three phonon processes.
The relaxation times from three-phonon scattering processes are given by (\ref{Gamma}).

\subsection{Tersoff Interatomic Potential}

In both our LD calculations and MD simulations the interaction between carbon atoms is described by the bond order empirical potential proposed by Tersoff~\cite{Tersoff1988, Tersoff1988a}.
The Tersoff potential has been designed to model covalently bonded systems and has been extensively applied to Si, C and Ge systems~\cite{Tersoff1989, Tersoff1990a}.
The original parameter set provided by Tersoff was developed to reproduce structural properties of bulk semiconductors such as silicon, amorphous carbon and diamond, and has been recently re-optimized to accurately reproduce the vibrational properties of sp$^2$ carbon nanostructures~\cite{Lindsay2010}.
In particular, this set of parameters provides a very good description of acoustic phonons, which are the major heat carriers in a material.

In the Tersoff formulation, the potential energy between atoms $i$ and $j$ can be written in the form
\begin{equation}
V_{ij} = f^C(r_{ij}) \left[ f^R(r_{ij}) + b(\theta_{ijk}) f^A(r_{ij}) \right],
\end{equation}
where $f^R(r_{ij})$ is a repulsive term, $f^A(r_{ij})$ is an attractive term and $b(\theta_{ijk})$ is a three body term which controls the strength of the attractive term depending on the number of neighbours $k$ of atoms $i$ and $j$, and on the angle between these atoms, $\theta_{ijk}$.  
The term $f^C(r_{ij})$ is a cutoff function which limits the range of atomic interactions in the system. Typically, the cutoff radius is such that only first neighbour interactions are considered.

Due to its bond order character, the Tersoff potential can accurately describe both sp$^2$ and sp$^3$ carbon systems, as well as different carbon structures with a single set of parameters.
Therefore, we apply the Tersoff potential in all simulations presented in this work, with the parameter set given in ~\cite{Lindsay2010}. \footnote{In the particular case of fcc-(C$_{28})_2$, the distance between second neighbours can be very close to the cutoff radius $R=0.21$ nm of the chosen parameter set. This issue can cause large peaks in the atomic forces during an MD simulation, which then requires a very small integration timestep. In order to keep the same timestep for all simulations, we shortened the cutoff radius to $R=0.19$ nm exclusively for the fcc-(C$_{28})_2$ systems. This change in cutoff radius does not affect any relevant physical property of the material.}

\subsection{Simulation Details}

All structures are generated with a crystalline atomic arrangement corresponding to the optimized configuration with periodic boundary conditions.
Each system is thermalized for $100$ ps with a Nos\'e-Hoover thermostat~\cite{Nose:1984wa} set at $300$ K while its volume is kept constant.
The system is them coupled to a barostat set at zero pressure and allowed to expand or contract the volume in order to achieve a relaxed configuration. 
This step depends on the system size and can last from $1$ ns to $10$ ns.
The relaxed structure is used as the starting configuration for every simulation performed afterwards, where the velocities are set to $300$ K and the system is again coupled to a Nos\'e-Hoover thermostat for $1$ ns in order to decorrelate the initial configurations.
Finally, the thermostat is decoupled from the system and the heat flux calculations are performed under microcanonical conditions, being recorded every $5$ MD steps.
During MD simulations the equations of motion are integrated with a velocity Verlet integrator with a timestep of $1$ fs.

In order to increase the accuracy of $\kappa$ at least $10$ independent configurations are averaged, and the uncertainty is taken as the standard deviation.
In the case of graphene and schwarzite, the final value of $\kappa$ is averaged over the equivalent diagonal components of the conductivity tensor. 
The total simulation time of each configuration determines the accuracy of the HFACF. For the SWCNT a total time of $120$ ns was required for each configuration. In the case of graphene, each simulation was at least $60$ ns long, and for the schwarzite structures at least $40$ ns.

Size convergence of $\kappa$ is a very important aspect of MD-based heat transport simulations, not only because of the size effects on fluctuations, but also because the phonon mean-free-path is limited by the size of the simulation cell.
Therefore, we simulate supercells of increasing size with periodic boundary conditions.
This is particularly important for low-dimensional systems with long wavelength phonons where  the thermal conductivity might diverge~\cite{Lippi2000, Lepri2003, Lepri2005, Wang2012}.
In effect, the convergence trend of $\kappa$ with system size provides an indirect estimate regarding the role of certain phonons in the heat transport process~\cite{Donadio2007, pereira2013}.

\section{Results}
\label{sec:results}

\subsection{Carbon Nanotubes}

Carbon nanotubes are excellent heat conductors. 
Regarding their electronic properties, SWCNTs can behave as metals or semiconductors, depending on their chirality. 
In semiconducting nanotubes, such as the (10,0) SWCNT, heat is mostly carried by phonons.
The high value of $\kappa$ found in CNTs is due to a combination of low dimensionality and a crystalline structure which favours long wavelength phonons~\cite{Mahan:2004fd}.

Whereas 1D momentum conserving systems present a power-law dependence of the thermal conductivity on the length~\cite{Lippi2000, Lepri2003, Lepri2005, Wang2012}, $\kappa \sim L^{\alpha}$, with $\alpha\sim$ 0.33 - 0.4, CNTs are not purely 1D and thus may have a large but finite $\kappa$. 
Although the size convergence of $\kappa$ for CNTs is still  debated~\cite{Savin:2009hr,Lindsay:2009cz}, anharmonic LD calculations and equilibrium MD simulations show that the conductivity of SWCNTs seems to converge with length in the limit of infinitely extended systems~\cite{Mingo2005, Donadio2007}. Divergent $\kappa$ and breakdown of Fourier's law for CNTs may be observed in the presence of large thermal fluxes~\cite{Zhang:2012ca}.

\begin{figure}[htbp]
\hfill
\includegraphics[width=0.75\linewidth]{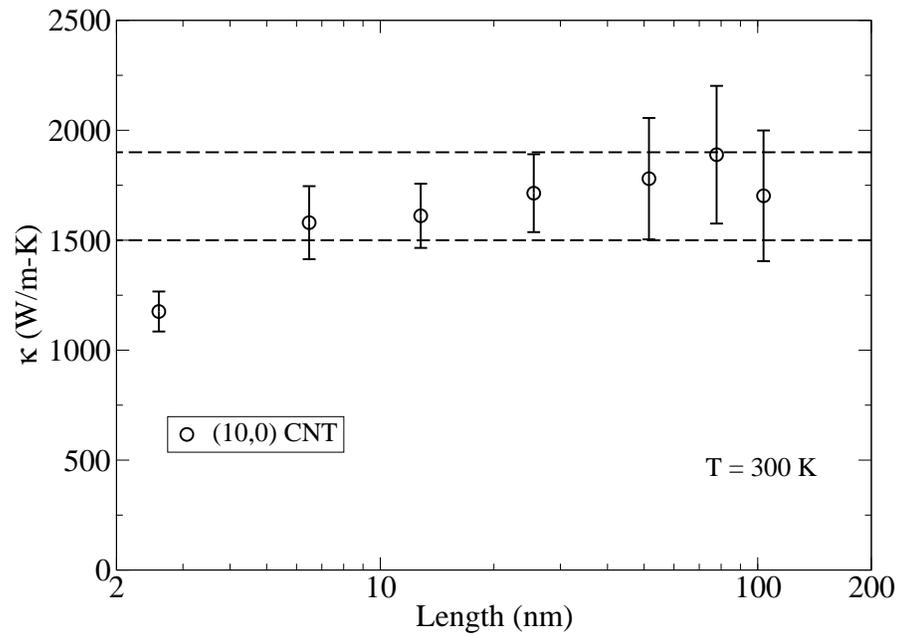}
\caption{Thermal conductivity of a (10,0) SWCNT as a function of the length of the supercell along the tube axis. The Green-Kubo calculation converges to $\kappa = 1700 \pm 200$ \wmk for $L>5$ nm. Dashed lines indicate the upper and lower limits given by the uncertainty in $\kappa$.}
\label{fig:gkswcnt}
\end{figure}

Using (\ref{eq:greenkubo}) we calculate $\kappa$ for SWCNT with supercells of increasing length along the nanotube axis as shown in figure \ref{fig:gkswcnt}.
For very short supercell length, there is poor sampling of low-frequency flexural modes and the thermal conductivity is underestimated. 
Another contributing factor to this underestimation is the excessive scattering of long wavelength phonons due to the finite size of the supercell.
Both limitations can be mitigated by increasing the length of the supercell in order to account for all relevant acoustic phonon modes appropriately.
The results converge to $\kappa = 1700 \pm 200$ \wmk for $L>5$ nm.
It is worth noting that this numerical value is in agreement with a previous work which employed the original parameter set provided by Tersoff, $\kappa = 1750 \pm 230$ \wmk ~\cite{Donadio2007, Donadio2009a}.

\subsection{Graphene}

This one atom thick 2D material presents remarkable physical properties which hold much promise to future technological applications.
Its very high thermal conductivity and long phonon mean free path rival those of nanotubes.
Similarly to CNTs, graphene is a low dimensional material and as such its thermal conductivity should, in principle, diverge with the sheet size.
In the particular case of 2D materials, the conductivity shows a logarithmic divergence, such that $\kappa \sim \log L$ ~\cite{Lippi2000, Lepri2003, Lepri2005, Wang2012}.

\begin{figure}[htbp]
\hfill
\includegraphics[width=0.75\linewidth]{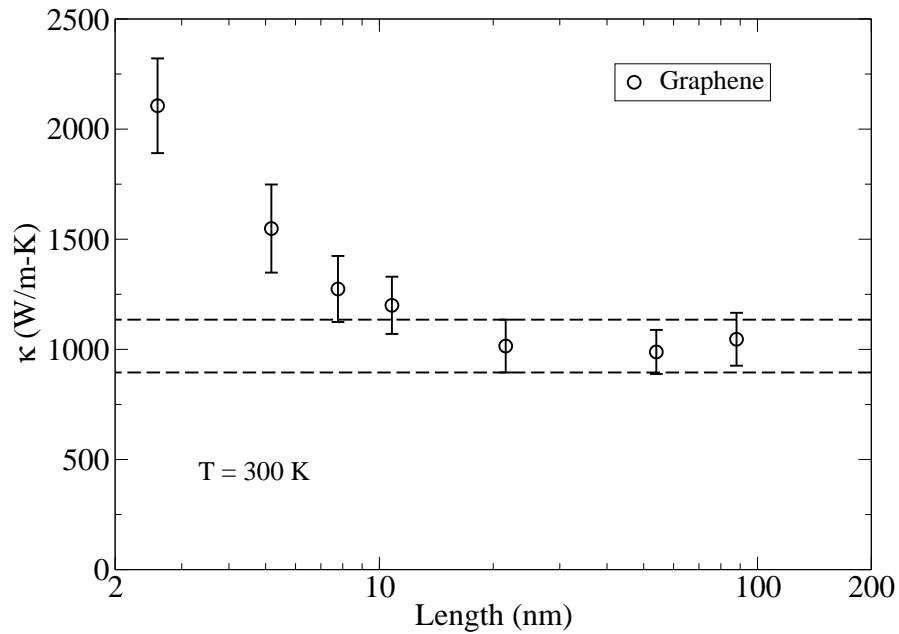}
\caption{Thermal conductivity of graphene as a function of supercell side length, for approximately square cells. Reported values are average over the two in-plane thermal conductivities. Converged value $\kappa = 1015 \pm 120$ \wmk is achieved for a cell with $\approx 20$ nm side length and $\approx 2 \cdot 10^4$ carbon atoms. Dashed lines indicate upper and lower limits given by the uncertainty.}  
\label{fig:gkgraphene}
\end{figure}

Nonetheless, by performing equilibrium MD simulations and calculating the  thermal conductivity by the Green-Kubo formula in (\ref{eq:greenkubo}), we observe a convergence of $\kappa$ with the size of the graphene supercell, as shown in figure \ref{fig:gkgraphene}. 
The reported conductivity values are calculated as the average over two components of the conductivity tensor, which are equivalent due to hexagonal lattice symmetry, such that $\kappa = (\kappa_{xx} + \kappa_{yy})/2$.
Each graphene supercell is approximately square ($L=L_x \approx L_y$), with the armchair edge oriented along the cartesian x-axis.
The smallest supercell has $\approx 200$ atoms and the largest $\approx 3\cdot 10^5$.
We obtain a converged value for the conductivity $\kappa = 1015 \pm 120$ \wmk, for a supercell with $\approx 2 \cdot 10^4$ carbon atoms~\cite{pereira2013}.
Similarly to the case of SWCNT, the size convergence of $\kappa$ for graphene has also been verified by anharmonic LD calculations~\cite{Bonini2012}. 

An interesting aspect of the convergence trend shown in figure \ref{fig:gkgraphene} is that, differently from the SWCNT case, $\kappa$ decreases as the supercell is increased.
Once again, as the supercell is increased we can probe longer wavelength phonons, which in graphene are typically acoustic flexural modes.
In direct contrast to SWCNT, the role of these flexural modes in graphene is to limit $\kappa$, avoiding the $\log$ divergence expected for 2D systems. 
We have also recently reported that in the absence of such flexural modes the conductivity of graphene diverges as expected~\cite{pereira2013}.

\subsection{fcc-(C$_{28})_2$ and fcc-(C$_{64})_2$ Schwarzites}

Starting from a single graphene sheet, it is possible to generate ``closed" (positively curved) $sp^2$ structures, such as buckyballs, by replacing  hexagonal rings by pentagons. Analogously, replacing hexagons by heptagons and octagons it is possible to generate open (3-dimensional negatively curved) $sp^2$ carbon crystalline structures~\cite{Mackay:1991tl}. Periodic negatively curved crystalline schwarzites have been predicted to be more stable than the  fullerenes with equal absolute curvature, however it has so far not been possible to synthesize them, because 7- and 8-membered rings tend to get stabilized by adjacency. The formation of abutting large rings promotes the growth of purely graphitic spongy carbon, which retains a highly ordered short-range structure but forms self-affine amorphous minimal surface~\cite{Donadio:1999ur, Bogana:2001wf,Barborini:2002dl}.   

Crystalline schwarzites can be of either octahedral P-type or tetrahedral D-type, made of units with six or four branches, respectively. P-type units form simple cubic periodic structures whereas D-type units arrange themselves in diamond structures. 
D-type schwarzites containing only 6 and 7-membered rings are particularly interesting, because it is possible to establish a one to one correspondence to fullerenes by replacing the 5-membered rings with 7-membered rings. The smallest schwarzite in this class is made of units of 28 carbon atoms forming solely 7-membered rings.  

To address heat transport in this class of materials we consider the smallest D-type schwarzite, (C$_{28})_2$, and a relatively large one, (C$_{64})_2$ made of 6- and 8-membered rings (Fig~\ref{fig:structures}c and d).
Stability, elastic properties and electronic structure of these systems were formerly characterized by semi-empirical and first-principles simulations~\cite{Gaito:1998ur,Spagnolatti:2003is}.
These previous works showed that the smallest schwarzites, like nanotubes, can be either metallic or semiconducting and that there is a close relation between topology and electronic properties.  Even though (C$_{28})_2$ and (C$_{64})_2$ are metallic, (C$_{28})_2$ displays very localized states, and both have a low density of electronic states at the Fermi level, which leads to poor electrical conductivity and electronic thermal conductivity.
In addition the electronic structure of larger systems with smaller Gaussian curvature would resemble that of rippled or distorted graphene, with an electronic energy gap at the Dirac point that directly depends on the curvature, thus suggesting the possibility for gap engineering.~\cite{Benedek:2010gv} 

\begin{figure}[htbp]
\hfill
\includegraphics[width=0.75\linewidth]{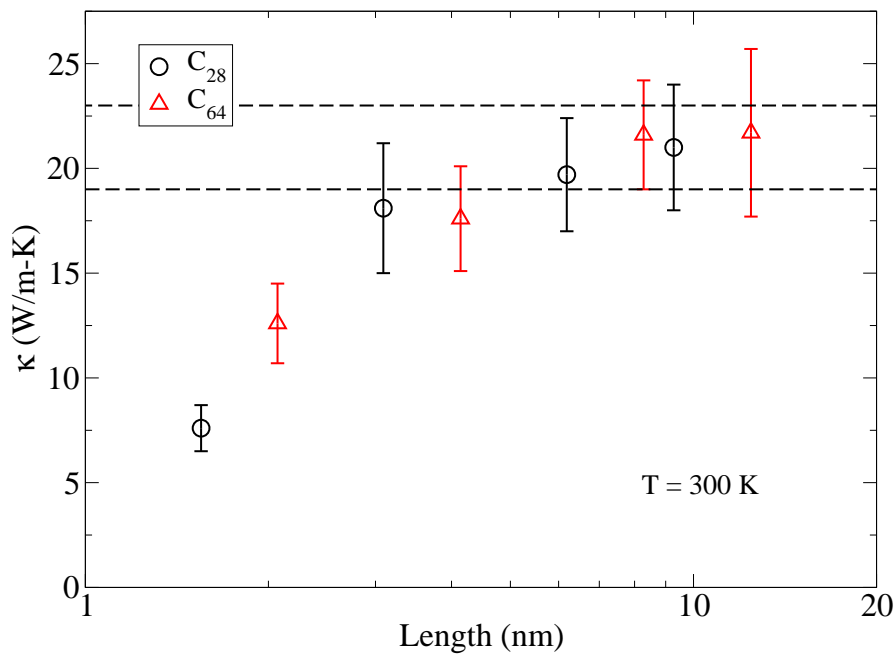}
\caption{Thermal conductivity of (C$_{28})_2$ and (C$_{64})_2$ schwarzites as a function of the side length for cubic supercells. Reported values are average the over the three equivalent components of the thermal conductivity. Converged value $\kappa = 21 \pm 2$ \wmk is achieved for a supercell with $\approx 8$ nm side length.}
\label{fig:gkschwarzites}
\end{figure}

As we did for SWCNT and graphene we have computed the phononic thermal conductivity of the D-type schwarzites (C$_{28})_2$ and (C$_{64})_2$ by equilibrium MD simulations of cubic supercells, in order to verify its size convergence.
Once again, $\kappa$ is calculated as the average over the three equivalent components of the conductivity tensor.
Our simulations show that $\kappa$ converges for supercells with linear dimension of about $8$ nm containing about 3$\cdot 10^4$ atoms. 
Surprisingly the bulk thermal conductivity of both schwarzites is $21 \pm 2$  \wmk, about 50 and 100 times lower than that of graphene and (10,0) SWCNT, respectively. 

The density of (C$_{28})_2$ and (C$_{64})_2$ schwarzites are $1.3$ and $1.2$ g/cm$^3$, i.e. $0.38$ and $0.34$ the density of diamond and about one half the density of graphene with a nominal thickness of $0.335$ nm. Since $\kappa$ is linearly proportional to the atomic density, these density ratios alone do not justify such a large reduction of the conductivity in schwarzites relative to other crystalline carbon structures.

\section{Discussion}
\label{sec:LD}

The schwarzites considered here are crystalline and periodic, there is neither mass disorder nor any other source of phonon scattering, and the type of chemical bonding is analogous to the other sp$^2$ forms of carbon considered. 
Therefore, the only possible reason for such a dramatic difference in the heat transport coefficient lies in substantial differences in the phonon properties at either harmonic or anharmonic level, which arise from the imposition of a negative curvature to graphene in order to form a 3D structure.
We have characterized the vibrational properties of (10,0) SWCNT, graphene and schwarzite computing  phonon dispersion curves and group velocities by harmonic LD. We have also calculated the phonon lifetimes by anharmonic LD as detailed in section~\ref{sec:methods}. 

\begin{figure}[htbp]
\hfill
\includegraphics[width=0.75\linewidth]{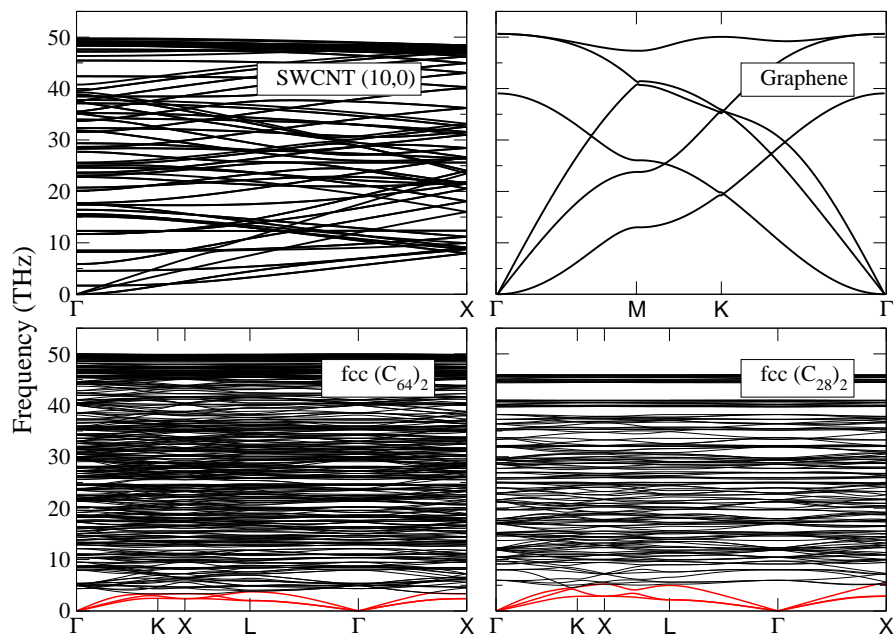}
\caption{Phonons dispersions for (10,0) SWCNT, graphene, and schwarzites fcc-(C$_{28})_2$ and fcc-(C$_{64})_2$, calculated within the harmonic approximation. In schwarzites, acoustic phonons, highlighted in red, are limited to the low-frequency region of the spectrum.}
\label{Fig:dispersions}
\end{figure}
Phonon dispersion curves are shown in figure~\ref{Fig:dispersions}. 
While graphene displays three acoustic and three optical branches, SWCNT and schwarzites have a large number of optical branches, stemming from the relatively large number of degrees of freedom in the unit cell.  
Nonetheless, the modes with frequency up to $\sim 45$ THz in SWCNT are very dispersive high order flexural modes~\cite{Mahan:2004fd}, whereas for schwarzites we observe that the optical modes have relatively flat dispersions, and limit the frequency range of the acoustic phonons between 0 and $\sim$7 THz. 
There is also a remarkable difference with respect to the acoustic modes of graphene, where the longitudinal acoustic branch extends beyond 40 THz.
\begin{figure}[htbp]
\hfill
\includegraphics[width=0.75\linewidth]{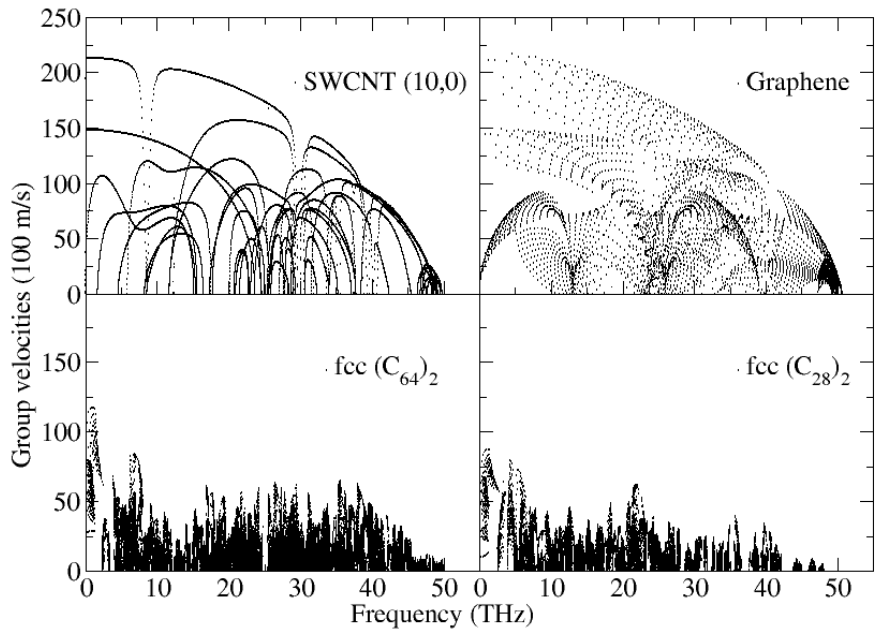}
\caption{Phonons group velocities for (10,0) SWCNT, graphene, and schwarzites fcc-(C$_{28})_2$ and fcc-(C$_{64})_2$, as a function of frequency calculated from phonon dispersions within the harmonic approximation. Phonons in schwarzite present much lower group velocities which accounts, at least in part, to the reduction in $\kappa$ as shown in (\ref{bte-smrt}).}
\label{Fig:velocities}
\end{figure}
These differences among phonon band structures imply large differences in the phonon group velocities of carbon materials of different dimensionality, as shown in figure~\ref{Fig:velocities}. Group velocities are much lower in schwarzites, when compared to SWCNT and graphene. 
Following (\ref{bte-smrt}), phonon group velocities contribute as a quadratic term to $\kappa$, therefore the observed suppression of $v_g(\omega,q)$ justifies, at least qualitatively, the reduction of $\kappa$ in schwarzites.

\begin{figure}[htbp]
\hfill
\includegraphics[width=0.75\linewidth]{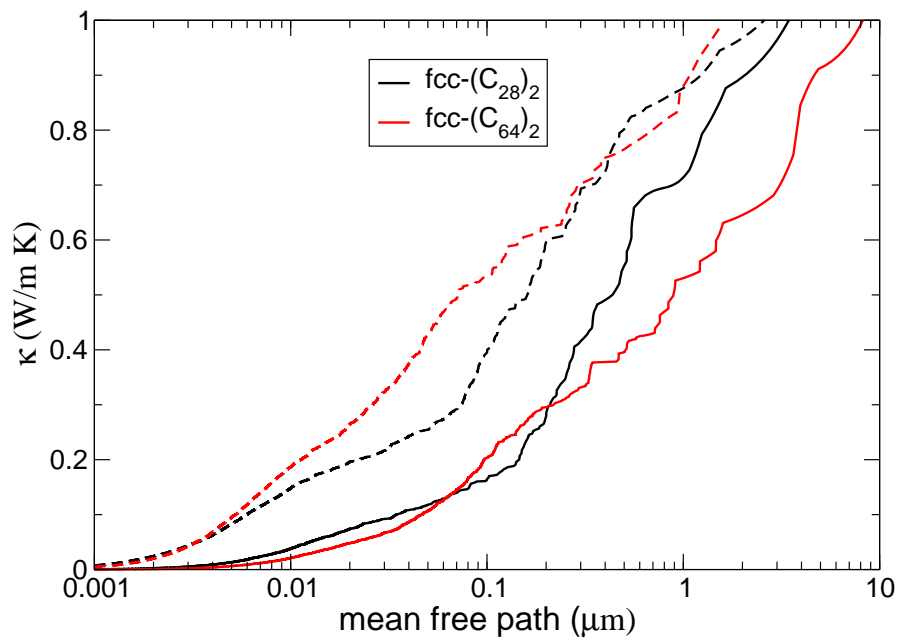}
\caption{Normalized accumulation functions for the thermal conductivity of fcc-(C$_{28})_2$ (black) and fcc-(C$_{64})_2$ (red), computed using quantum (solid lines) and classical statistics (dashed lines).}
\label{Fig:accumulation}
\end{figure}
To get a quantitative picture of the specific phonon contributions to the thermal conductivity of schwarzites, we have computed the phonon lifetimes of (C$_{28})_2$ and (C$_{64})_2$ by evaluating (\ref{Gamma}) for the conventional cubic cells using 6x6x6 and 4x4x4 shifted q-point meshes~\cite{MONKHORST:1976ta,Savic:2013tg}. 
The low frequency phonon lifetimes of the schwarzites are larger than those of the flexural modes of graphene, but decrease faster ($\tau\propto\omega^{-1.8}$), so that there is a cross-over at relatively low frequency $\sim 5$ THz, after which the phonon modes of graphene not only have much larger group velocities, but also larger relaxation times. It is worth noting that the lifetimes of both the longitudinal and the transverse acoustic modes of graphene are nearly frequency independent~\cite{Bonini2012}, so that they also contribute to amplify the difference in $\kappa$ between graphene and schwarzites.

The total thermal conductivity from anharmonic LD is given by (\ref{bte-smrt}). The possibility to perform anharmonic LD calculations using either classical or quantum phonon distribution functions permit the evaluation of the errors occurring in MD simulations from neglecting nuclear quantum effects.  While we get a very good agreement between MD and anharmonic LD with classical statistics,  using the correct quantum distribution functions we get about twice as large $\kappa$ for both schwarzites. For (C$_{28})_2$ $\kappa_{cl}=28$ \wmk and $\kappa_{QM}=57$ \wmk . For (C$_{64})_2$ $\kappa_{cl}=26$ \wmk and $\kappa_{QM}=64$ \wmk . 
The comparison between quantum and classical normalized accumulation functions (figure \ref{Fig:accumulation}), i.e. the cumulative contribution to $\kappa$ as a function of the mean free path, shows that quantum effects largely enhance the contribution of phonons with micrometer long mean free path. These are typically low-frequency phonons with wavelength of several tens of nanometers, therefore their contribution may be attenuated by further disorder at the same length-scale. 

\section{Conclusions}
\label{sec:conclusions}

In conclusion, we have shown that the thermal conductivity of two D-type schwarzites converge with supercell size to a value $\kappa = 21 \pm 2$ \wmk. 
In spite of its carbon sp$^2$ bonded crystalline structure, the schwarzites have a thermal conductivity about 50 and 80 times lower than SWCNT and graphene, calculated by classical molecular dynamics at equilibrium with the same empirical potential.
This remarkable difference between the thermal conductivities of similarly sp$^2$-bonded carbon materials cannot be explained solely by the difference in atomic density among these materials. The reduction of $\kappa$ stems from the change in dimensionality from low-D to 3D, which drastically changes the nature of the phonon modes in the system. The absence of flexural modes and the large number of atoms in the  unit cells of schwarzites limits the frequency range of the acoustic modes and greatly reduces their group velocities. 
Nevertheless, the accumulation functions show that phonons with long mean free path larger than 1 $\mu$m provide the major contribution to heat transport. 

A 3D all-carbon material such as schwarzite, with such low thermal conductivity and its corresponding electronic properties might provide the route for high efficiency carbon-based thermoelectrics.
In fact the observed reduction of $\kappa$ from graphene to schwarzites is comparable to that that can be achieved by chemical functionalisation of graphene~\cite{Kim:2012gn}, but schwarzites present the advantage of being stable three-dimensional structures. In addition the thermal conductivity of schwarzites may be further reduced by intercalation, or naturally occurring mesoscale disorder, which may contribute to scatter longer mean free path phonons.

\ack{We are grateful to G. Benedek and M. Bernasconi for providing the schwarzite models and for useful discussions, and to Denis Andrienko for a critical reading of the manuscript.
We acknowledge the provision of computational facilities and support by Rechenzentrum Garching of the Max Planck society (MPG), and access to the supercomputer JUGENE at the J\"ulich Supercomputing Centre under NIC project HMZ26. 
Financial support was provided by MPG under the MPRG program.}

\section*{References}
\bibliography{library,morelibrary}
\bibliographystyle{iopart-num}

\end{document}